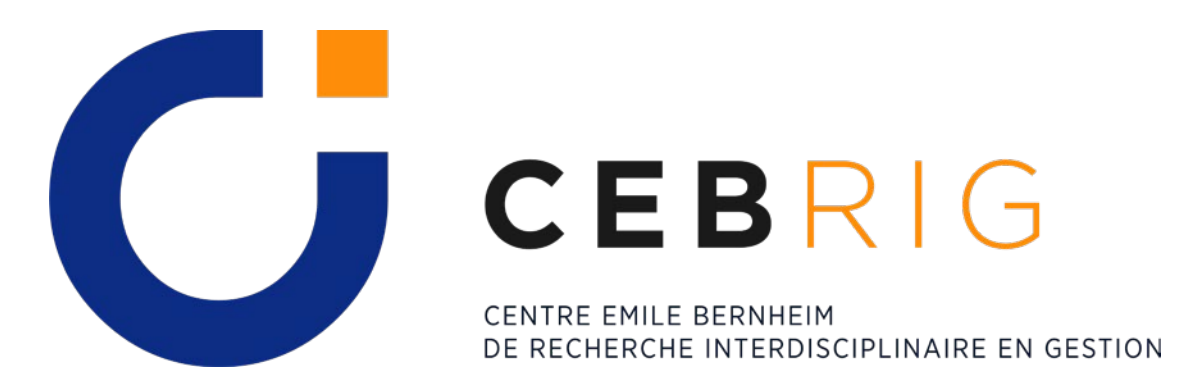

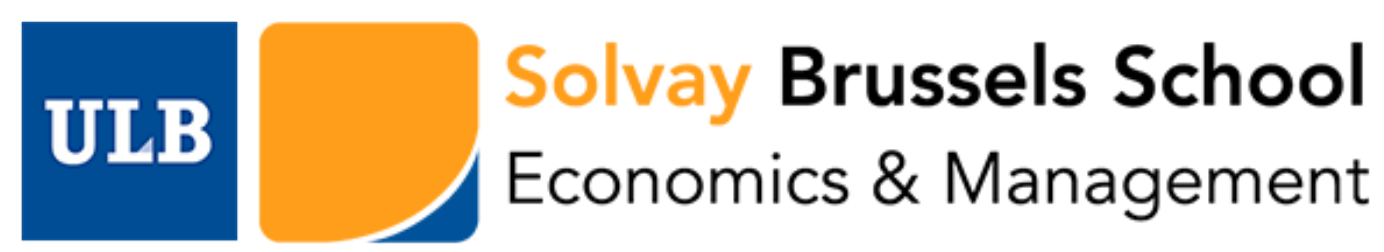

# Developing Bayesian probabilistic reasoning capacity in HSS disciplines: Qualitative evaluation on bayesvl and BMF analytics for ECRs

Quan-Hoang Vuong & Minh-Hoang Nguyen

Methodological innovations have become increasingly critical in the humanities and social sciences (HSS) as researchers confront complex, nonlinear, and rapidly evolving socio-environmental systems. On the other hand, Early Career Researchers (ECRs) continue to face intensified publication pressure, limited resources, and persistent methodological barriers. Employing the GITT–VT analytical paradigm—which integrates worldviews from quantum physics, mathematical logic, and information theory—this study examines the seven-year evolution of the Bayesian Mindsponge Framework (BMF) analytics and the bayesvl R software (hereafter referred to collectively as BMF analytics) and evaluates their contributions to strengthening ECRs' capacity for rigorous and innovative research. Since 2019, the bayesvl R package and BMF analytics have supported more than 160 authors from 22 countries in producing 112 peer-reviewed publications spanning both qualitative and quantitative designs across diverse interdisciplinary domains. By tracing the method's inception, refinement, and developmental trajectory, this study elucidates how accessible, theory-driven computational tools can lower barriers to advanced quantitative analysis, foster a more inclusive methodological ecosystem—particularly for ECRs in low-resource settings—and inform the design of next-generation research methods that are flexible, reproducible, conceptually justified, and well-suited to interdisciplinary inquiries.

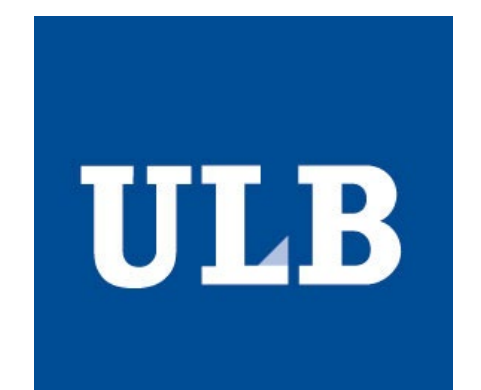

*Université Libre de Bruxelles - Solvay Brussels School of Economics and Management*
*Centre Emile Bernheim de Recherche Interdisciplinaire en Gestion*
*ULB CP114/03 50, avenue F.D. Roosevelt 1050 Brussels BELGIUM*
*cebrig@ulb.be - Tel : +32 (0)2/650.48.64*

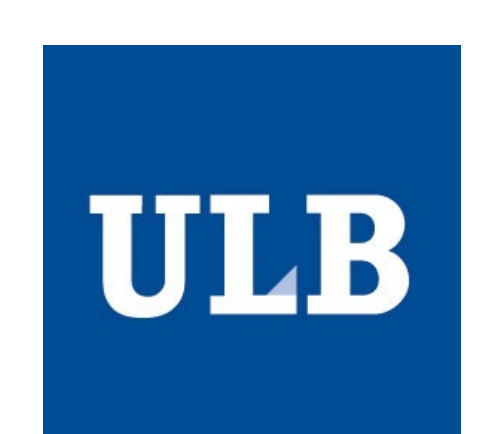

**Developing Bayesian probabilistic reasoning capacity in HSS disciplines: Qualitative evaluation on bayesvl and BMF analytics for ECRs**

Quan-Hoang Vuong [1,2,3], Minh-Hoang Nguyen [1,*]

[1.] Centre for Interdisciplinary Social Research, Phenikaa University, Hanoi, Vietnam

[2.] Professor. University College, Korea University. 145 Anam-ro, Seongbuk-Gu, Seoul 02841, Republic of Korea

[3.] CEBRIG, Université Libre de Bruxelles, Brussels 1050, Belgium

*** Correspondence:** hoang.nguyenminh@phenikaa-uni.edu.vn

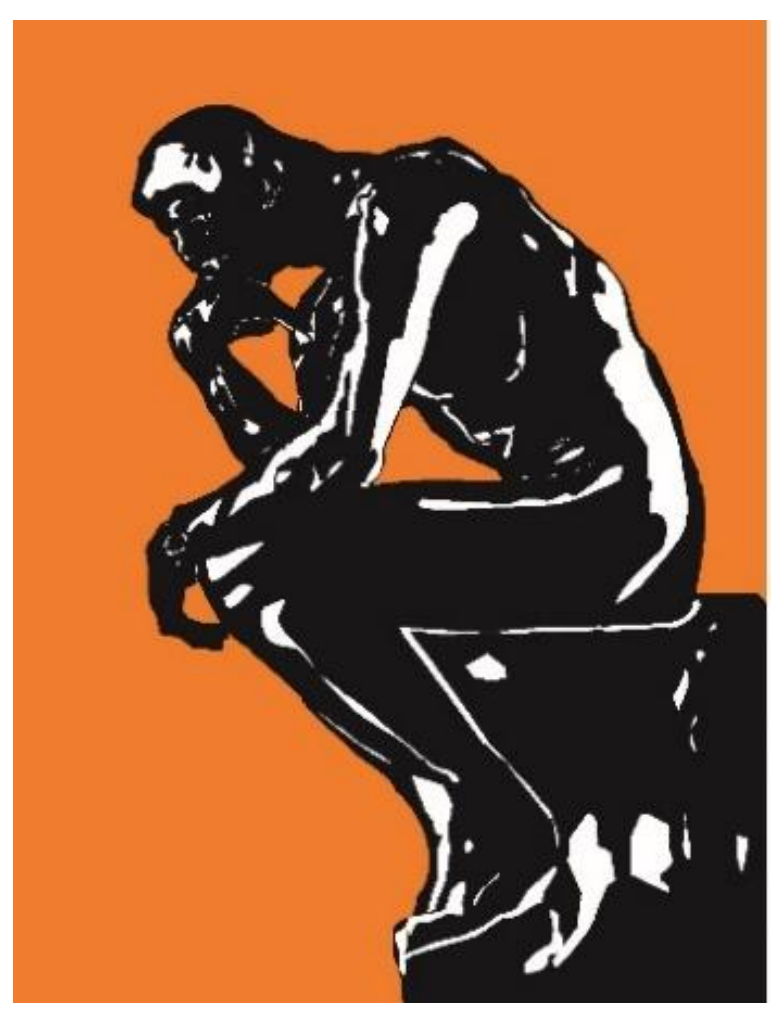

"At a high level of knowledge, learning naturally has to be paired with practice."

– In "Bird Village Economics," *Wild Wise Weird* (2024)

## Abstract

Methodological innovations have become increasingly critical in the humanities and social sciences (HSS) as researchers confront complex, nonlinear, and rapidly evolving socio-environmental systems. On the other hand, Early Career Researchers (ECRs) continue to face intensified publication pressure, limited resources, and persistent methodological barriers. Employing the GITT–VT analytical paradigm—which integrates worldviews from quantum physics, mathematical logic, and information theory—this study examines the seven-year evolution of the Bayesian Mindsponge Framework (BMF) analytics and the bayesvl R software (hereafter referred to collectively as BMF analytics) and evaluates their contributions to strengthening ECRs' capacity for rigorous and innovative research. Since 2019, the bayesvl R package and BMF analytics have supported more than 160 authors from 22 countries in producing 112 peer-reviewed publications spanning both qualitative and quantitative designs across diverse interdisciplinary domains. By tracing the method's inception, refinement, and developmental trajectory, this study elucidates how accessible, theory-driven computational tools can lower barriers to advanced quantitative analysis, foster a more inclusive methodological ecosystem—particularly for ECRs in low-resource settings—and inform the design of next-generation research methods that are flexible, reproducible, conceptually justified, and well-suited to interdisciplinary inquiries.

## 1. Introduction

The world is confronting a series of interconnected crises that threaten both societal stability and the planet's long-term viability. Climate change and environmental degradation have pushed the Earth beyond several safe planetary boundaries, increasing the likelihood of irreversible, cascading disruptions that imperil human survival and development (Armstrong McKay et al., 2022; Richardson et al., 2023). Simultaneously, widening socioeconomic inequalities fuel social unrest, political polarization, and radicalization (Franc & Pavlović, 2023; Qureshi, 2023). Adding to this turbulence is the rapid advancement of emerging technologies—particularly artificial intelligence (AI)—whose unchecked development poses risks ranging from economic displacement and misinformation to extreme power concentration and even existential threats (Marr, 2023; Turchin & Denkenberger, 2020). In such a volatile context, methodological innovation becomes indispensable for the social sciences and humanities, which must analyze complex, fast-moving social, cultural, and environmental dynamics that traditional approaches struggle to capture.

New research methods are essential not only to foster deeper interdisciplinary research but also to integrate advanced computational capabilities. Bayesian inference is one such approach. Scholars have long argued that Bayesian analysis offers a more coherent and unified framework for statistical reasoning in the social sciences (Gill, 2015; Jackman, 2000, 2009). By treating all unknown quantities probabilistically, Bayesian methods encourage greater caution and prudence in evaluating evidence, helping researchers avoid exaggerated effect claims and logical inconsistencies (Gill, 2015; Kruschke, 2014; McElreath, 2018). Because social science data often unfold over time and rarely meet assumptions of infinite sampling, the Bayesian capacity to update priors with newly observed information is especially advantageous, allowing researchers to explicitly refine inferences and account for estimation biases (Gill, 2015).

Bayesian inference also addresses several limitations of conventional frequentist approaches, which have contributed to the reproducibility crisis in socio-psychological research (Camerer et al., 2018; Open Science Collaboration, 2015). Problems such as fickle *p*-values, their widespread misinterpretation, and the arbitrary use of significance thresholds (Halsey, Curran-Everett, Vowler, & Drummond, 2015) diminish the interpretability and robustness of findings. Although estimation-focused approaches using means and confidence intervals offer improvements (Masson & Loftus, 2003), Bayesian inference provides a fully integrated probabilistic framework that unifies estimation, uncertainty quantification, and visualization. This enhances transparency and conceptual coherence—an essential requirement in socio-behavioral research, where contextual uncertainty and cognitive complexity shape theoretical reasoning.

Despite these advantages and the rise of computational tools such as Markov chain Monte Carlo (MCMC), Bayesian methods remain underutilized in the social sciences and humanities. Persistent psychological barriers—including fear of mathematical complexity, fear of coding, and fear of leaving methodological comfort zones—continue to impede adoption. These barriers are especially acute for Early Career Researchers (ECRs), who face intense "publish or perish" pressures while often lacking adequate mentorship, resources, and methodological training (Callier & Polka, 2015; El Halabi et al., 2021). Under such constraints, ECRs tend to rely on familiar methods rather than explore innovative analytical tools. This situation underscores the need for methodological innovations that are both conceptually rigorous and user-friendly—tools that lower the entry barrier while supporting high-quality research.

In this context, the Bayesian Mindsponge Framework (BMF) analytics and the bayesvl software—built and running on R and Stan—represent a serious methodological attempt aimed at making Bayesian inference more accessible and pedagogically effective for interdisciplinary research in the social sciences and humanities (Vuong & La, 2025; Vuong, Nguyen, & La, 2022). BMF analytics integrates the reasoning capacity of Mindsponge Theory with Bayesian statistical inference, offering a coherent and innovative approach to scientific investigation and manuscript development. Complementing this framework, the bayesvl package—an open-source extension for R—supports Bayesian multilevel modeling using Stan and MCMC techniques. Built upon ggplot2 and rstan, and inspired by Bayesian networks and directed acyclic graphs (DAGs), bayesvl allows researchers to automatically construct regression models and generate Stan code from intuitive "relationship trees," thereby improving usability, transparency, and pedagogical effectiveness (La et al., 2022). Because bayesvl functions as an essential component of BMF analytics, we refer to both collectively as BMF analytics throughout the remainder of this paper.

After seven years of continuous development since 2019, the methods have enabled more than 160 authors from 22 countries to publish 112 peer-reviewed studies. These works span both qualitative and quantitative designs and cover a wide array of interdisciplinary topics, including environmental psychology, digital gaming, entrepreneurship, education, public health, cultural evolution, and research practices. Given the demonstrated usefulness and versatility of BMF analytics, examining its inception, refinement, and developmental trajectory offers important insights for future methodological innovation in the social sciences and humanities. First, such an investigation can illuminate how accessible, theory-driven computational tools help lower barriers to advanced quantitative research, thereby fostering a more inclusive methodological ecosystem—particularly for Early Career Researchers (ECRs) and scholars in low-resource environments. Second, understanding the empirical and pedagogical impacts of BMF analytics can inform strategies for designing next-generation research methods that are flexible, reproducible, conceptually grounded, and well-suited for interdisciplinary inquiry.

This study aims to examine the seven-year evolution of BMF analytics and evaluate their contributions to Early Career Researchers' capacity for rigorous and innovative research. The analysis was conducted using the GITT–VT analytical paradigm that integrates Granular Interaction Thinking Theory (GITT), information theory, mathematical logic, and worldviews from quantum mechanics.

## 2. GITT–VT Analytical Paradigm

The GITT–VT analytical paradigm represents an advanced methodological extension of the Bayesian Mindsponge Framework (BMF) analytics (Vuong & Nguyen, 2024b; Vuong, Nguyen, et al., 2022). This development emerges from expanding Mindsponge Theory—the theoretical foundation of BMF—into the broader Granular Interaction Thinking Theory (GITT) by integrating worldviews drawn from quantum physics and Shannon's information theory (Hertog, 2023; Rovelli, 2018; Shannon, 1948). Through this integration, the paradigm strengthens mathematical logic, enhances theoretical coherence and rigor, and enables scholars to construct self-contained qualitative theoretical or conceptual studies without requiring statistical modeling. In its full form, the paradigm consists of two complementary components: (1) qualitative analysis using the GITT lens and (2) quantitative analysis through BMF analytics. In the present study, we employed only the qualitative GITT component.

This methodological choice is appropriate for two main reasons. First, the paradigm is specifically designed to handle complexity, nonlinearity, and system dynamics—features that characterize interdisciplinary theorizing in socio-technological, ecological, and economic systems. It therefore offers a suitable foundation for examining the dynamic, nonlinear processes (e.g., serendipitous moments) that have shaped the emergence and evolution of BMF analytics. Second, as an extension of Mindsponge Theory, GITT preserves the theory's core strengths in explaining creativity and innovation formation—capabilities that are central to the Serendipity-Mindsponge-3D (SM3D) knowledge-management framework. In this sense, the SM3D framework can be viewed as embedded within GITT (Nguyen, Jin, et al., 2023; Vuong, Le, et al., 2022). This framework has been employed to analyze innovation and creativity across diverse domains, including firm-level innovation (Sun, Ruan, Peng, & Lu, 2022), urban technological advancement (Jiang & Xiong, 2024), Covid-19 vaccine development (Vuong, Le, et al., 2022), digital creativity (Nguyen, Jin, et al., 2023), ethical innovation (Heydarkhani, Khamseh, & Kheradranjbar, 2024), and dataset construction and disclosure (Tian, Cheng, Xue, Han, & Shan, 2023).

Grounded in quantum mechanics and information theory, GITT posits that observable phenomena arise from interactions among finite informational quanta. Under this view, reality is fundamentally granular, relational, and indeterminate (Rovelli, 2018):

- Granularity: Information within any system—including the human mind—is finite.
- Relationality: All events occur through interactions among systems; psychological processes reflect interactions between existing mental information and newly absorbed environmental inputs.
- Indeterminacy: The future is probabilistically shaped rather than strictly determined by the past, rendering psychological outcomes—including knowledge creation—inherently probabilistic.

GITT comprises two primary interacting spectrums: the mind and the environment. The mind functions as an information-absorbing, -processing, and -generating system, while the environment—which includes socio-cultural structures, economic systems, ecosystems, and the planetary climate—constitutes a broader informational landscape (Vuong, La, & Nguyen, 2025b). Because all events are interactional and all properties exist only relationally, the mind is in continuous reciprocal exchange with its surroundings, reorganizing itself to maintain coherence and viability. Systems that regulate these informational exchanges effectively are those that adapt, persist, and flourish. Put differently, survival and evolution in a dynamic world depend on the efficient management of information—its acquisition, storage, transmission, and processing—a principle that resonates with Darwin's emphasis on adaptation as the foundation of survival (Darwin, 1964; Darwin & Wallace, 1858).

Following this logic, innovation can be understood as a novel change that is useful for the survival, development, and reproduction of a system. Such novelty emerges not only from interactions among information, experiences, knowledge, beliefs, and abilities within the mind, but also from informational inputs absorbed from the environment. Through these interactions, informational units are distinguished, evaluated, connected, compared, and transformed into raw materials for imagination to generate potentially useful insights. Importantly, perceptions of "novelty" and "usefulness" are relative to the pre-existing informational structure of the mind; a change need not possess objective novelty or usefulness to be internally recognized as innovative (Nguyen, Jin, et al., 2023).

While a larger pool of informational units provides greater potential for generating new ideas, it simultaneously increases the mind's entropy—its uncertainty and unpredictability. Shannon (1948)'s entropy formula formalizes this:

$$H(X_t) = -\sum_{i=1}^{n} P\big(x_i(t)\big) \log_2 P\big(x_i(t)\big)$$

where $H(X_t)$ denotes the entropy of a mental system $X$ at time $t$ with possible cognitive states $\{x_1, x_2, \dots, x_n\}$ and corresponding probabilities $\{P(x_1), P(x_2), \dots, P(x_n)\}$. Each probability $P(x_i)$ reflects the likelihood of the state $x_i$ occurring at time $t$. Within this framework, Entropy rises rapidly when informational units accumulate without prioritization, reaching its maximum when all cognitive states are equally probable—i.e., when $P\big(x_i(t)\big) = \frac{1}{n}$.

Therefore, a multi-filtering system is essential for reducing $H(X_t)$, enhancing internal coherence, and conserving cognitive energy. Such filtering reduces entropy either by concentrating probability on specific innovative elements or by removing irrelevant or costly information, thereby producing greater order and structure. For this system to function effectively, benchmarks—such as goals, philosophies, principles, priorities, and standards—are required to guide selective information processing.

The innovation process does not end when an idea is generated and enacted. Once an innovative product, service, or problem-solving action interacts with reality—or with other people—it generates feedback. This feedback introduces new information that may confirm, challenge, or refine the original idea. Because mental information is inherently subjective, shaped by individual experiences and biases, it cannot entirely mirror objective reality. The resulting discrepancy between internal representations and external conditions increases cognitive entropy.

To reduce this entropy and enhance the reliability of innovation, the mind must continuously distinguish, evaluate, compare, connect, and imagine. This updating process is formally captured by Bayes's theorem:

$$P(\,Ino \mid I_t\,) = \frac{P(\,I_t \mid Ino\,)P(Ino)}{\sum P(\,I_t \mid Ino_i\,)P(Ino_i)}$$

Here:

- $P(\,Ino \mid I_t\,)$ represents the updated or refined innovation after incorporating the new information.
- $P(Ino)$ is the prior—the mind's initial belief about the innovation before receiving new information.
- $I_t$ denotes the new information or feedback received at time $t$.
- $P(\,I_t \mid Ino\,)$ is the likelihood of observing the feedback under the current innovation.

This formulation illustrates how the mind systematically revises innovation based on accumulated evidence and knowledge. Innovation, therefore, is not a singular event but a continuous Bayesian updating cycle in which each interaction with the world provides informational inputs that refine the idea, reduce entropy, and increase its alignment with reality. This disciplined process must be repeated until the innovation adequately addresses real-world problems.

During this process, some information may initially appear trivial or useless. Such "perceived uselessness" may stem either from triviality (the information seems too obvious) or from perceived insignificance (its impact appears negligible). Kolmogorov's Zero–One Law helps illuminate this. If the $\sigma$-algebras $A_n$ are independent and $A$ is any event belonging to the tail σ-algebra, then $P(A)$ is either 0 or 1—implying triviality. In GITT terms, information that appears persistently unused may seem trivially obvious simply because it has never demonstrated utility under previous cognitive configurations. However, when the information structure within the mind reorganizes due to new inputs, previously trivial information may acquire new significance and become crucial for improving the innovation. This nonlinear process exemplifies serendipity, wherein the mind detects and leverages unexpected information for survival and development. The "unexpectedness" of such information can be tied to its trivial or seemingly insignificant nature (Vuong, 2022; Vuong, La, & Nguyen, 2025c).

Overall, an innovation outcome is a novel and useful outcome that emerges from a dynamic, multi-state, entropy-reduction knowledge management process undertaken by an individual or group to address issues critical to their survival, development, and reproduction.

Although the process requires disciplined management, serendipitous moments frequently occur and serve as catalytic inputs for innovation generation.

In the next section, we apply this GITT-based conceptualization of the innovation-making process to examine and elucidate the inception, refinement, and development of BMF analytics, as well as its impacts on early-career researchers.

## 3. The BMF analytics and bayesvl R package

The initial seed of the BMF analytics was the bayesvl software (version 0.8.5), which was published on CRAN on the evening of May 24, 2019 (La & Vuong, 2019). By 2022, a more comprehensive scientific computing approach—grounded in reflective analysis of successful scientific outputs that had applied bayesvl—had taken shape. This maturation culminated in the publication of a methodological book on Bayesian computational simulation, commonly referred to as BMF analytics, released by Sciendo / Walter de Gruyter GmbH on October 10, 2022 (Vuong, Nguyen, et al., 2022). As of today, the method has entered its seventh year of existence.

Although bayesvl and BMF analytics can be viewed as an integrated whole, BMF analytics is not technically restricted to bayesvl; it can be implemented using other Markov chain Monte Carlo (MCMC) engines or Bayesian computational tools. Nevertheless, employing BMF analytics in conjunction with bayesvl offers efficiency advantages, as both were designed according to a shared philosophy, priorities, and guiding principles. Conversely, bayesvl can also be used independently of BMF analytics. The software has since been upgraded to version 1.0.0—released on GitHub in August 2022 and on CRAN in May 2025—to ensure long-term stability, compatibility, and ease of updating (Vuong & La, 2025).

Drawing on GITT's conceptualization, the formation and evolution of BMF analytics can be divided into four main components. The first concerns the existential motivations underlying the demand for an interdisciplinary scientific research methodology. The second involves the articulation of philosophy, principles, and priorities to guide methodological development and reduce uncertainty. The third highlights methodological innovation as a dynamic, multi-state process of entropy reduction. The fourth examines the impact of BMF analytics on ECRs, particularly through initiatives such as the SM3D Portal's community coaching method, which was designed to help ECRs and researchers in low-resource settings overcome structural inequalities and initiate sustainable research careers, as well as the growing need to apply BMF analytics to ECRs' research objectives.

### 3.1. Inception from survival and development demand

The decision to develop bayesvl, and subsequently BMF analytics, emerged from the practical demands of work and institutional survival at the Centre for Interdisciplinary Social Research (ISR), Phenikaa University. The earliest ideas can be traced back to late 2017, shortly after ISR was established. The center's initial "start-up" years—if the term may be used—were marked by persistent uncertainty and risk. Financial resources were limited, workloads were heavy, and human capital was scarce: only four members, two of whom were full-time staff.

At that time, conducting research and publishing internationally remained largely unfamiliar to most lecturers and researchers in Vietnam, with the exception of a small number of long-established and well-resourced centers, primarily in the natural sciences and engineering. To sustain ISR's existence, a fundamental question continuously demanded an answer: Could the center maintain its scientific productivity? If so, how—by what means, with which tools, and on what timeline?

In seeking solutions, ISR members mobilized their full intellectual capacity, continuously observing, updating, and comparing available information on scientific research methodologies. At that juncture, researchers faced difficult choices. Classical frequentist statistical approaches appeared to offer certain practical advantages; newly produced frequentist-based studies from the center continued to be accepted, while earlier publications were beginning to accumulate impact. A shift toward Bayesian statistics, therefore, entailed great opportunity costs.

Despite this, ISR chose the Bayesian path and committed to developing bayesvl. One key advantage the research team identified was that Bayesian statistics provide a fertile intellectual space for reasoning, interpretation, and argumentation. At the time, the team believed that such conceptual openness was essential to meeting their survival needs.

Similar to bayesvl, the emergence of BMF analytics as a research and scientific writing methodology was likewise driven by survival imperatives. The idea of BMF analytics took shape under exceptionally challenging circumstances, as the world confronted the COVID-19 pandemic. The scientific community—particularly in the social sciences and humanities—was severely affected: publication processes in non-COVID-related fields became congested and stagnant. After 2020, the number of retracted articles increased rapidly, intensifying concerns over the replication crisis in psychology and social sciences, as well as transparency in research conduct and reporting.

In Vietnam, Hanoi underwent strict social distancing measures; schools and businesses were closed, and mobility was heavily restricted. Discussions about developing BMF analytics often took place on stone benches near researchers' homes. At the same time, ISR anticipated that post-pandemic socio-economic recovery pressures would significantly reduce research funding while substantially increasing research costs.

In this climate of heightened risk and uncertainty, ISR recognized the urgent need to further enhance labor productivity, reduce research costs, and increase methodological reliability. These contextual pressures and demands created the conditions for the birth of BMF analytics.

In practice, the emergence of BMF analytics was also the result of serendipity. In late 2020, the ISR team initiated a book project focused on applications of the Mindsponge mechanism—later fully developed into Mindsponge Theory in 2023 and Granular Interaction Thinking Theory (GITT) in 2024. During this process, a team member posed a pivotal question: Could Bayesian quantitative methods help overcome the limitations of frequentist statistics in studying the emergence of suicidal ideation—a complex, multidimensional phenomenon shaped by interactions among multiple factors?

This question revealed the deep compatibility and potential synergy between the Mindsponge mechanism and Bayesian analysis in social science and psychological research. Consequently, the first scientific output employing what would later be termed BMF analytics was the article "Alice in Suicideland: Exploring the Suicidal Ideation Mechanism through the Sense of Connectedness and Help-Seeking Behaviors," published on April 1, 2021, in *IJERPH*. However, the term BMF analytics was officially introduced only later, alongside bayesvl, at the Applied Statistics Seminar (August 2021) organized online by the Vietnam Institute for Advanced Study in Mathematics (VIASM) on August 4, 2021. This event also marked the launch of the book "*Bản hòa tấu dữ liệu xã hội*" ["*The Symphony of Social Data*"], a practical guide to data analysis using bayesvl, to the Vietnamese research community.

The first academic article to explicitly mention BMF analytics was “The roles of female involvement and risk aversion in open access publishing patterns in Vietnamese social sciences and humanities,” published on December 11, 2021, in the *Journal of Data and Information Science*. Finally, on October 10, 2022, the book “*The Mindsponge and BMF Analytics for Innovative Thinking in Social Sciences and Humanities”* was officially published by Sciendo / Walter de Gruyter GmbH, providing a systematic and intuitive introduction to BMF analytics. The release of bayesvl version 1.0.0 on GitHub on May 18, 2022—presented in Chapter 10 of the book—further solidified its role as a direct technical support tool for implementing the BMF analytics methodology.

### 3.2. Philosophy and principles

Developing a new methodology is not merely a matter of coding; it also involves deliberate design and careful consideration of how the method will be used to address specific scientific problems. Recognizing this, in parallel with deepening the knowledge of mathematics and computing and referring to existing scientific methods, researchers at ISR continuously engaged in collective discussions to distill a set of guiding philosophies and principles. These served to orient development and reduce risks throughout the methodological design process. In other words, they helped ensure that, once completed, the method would be useful for real research work.

Specifically, ISR has consistently upheld five core philosophies and principles in the design and development of the method.

First, the tool must support the reduction of financial costs in the research process (Vuong, 2018). Guided by a cost-effectiveness and efficiency principle, ISR emphasized open documentation and community support when developing bayesvl, enabling researchers—especially ECRs and those working under resource constraints—to easily access and apply the software. The first user manual for bayesvl (v0.8.1) was uploaded to the Open Science Framework on May 17, 2019, one day before the software was submitted to CRAN. Following its official release, usage guidelines and a condensed version of the BMF analytics methodology were published as open-access articles in *MethodsX* (Nguyen, La, Le, & Vuong, 2022; Vuong, La, et al., 2020) and *SoftwareX* (La et al., 2022).

Second, the research team actively promotes a culture of openness with respect to data, code, and early-stage results (Vuong, 2020). This culture enhances transparency, reduces risks arising from unexpected errors, and contributes to lowering research costs in the long run. When applying BMF analytics, authors are encouraged to upload their code and data to online repositories such as OSF and Zenodo and to report them transparently at the end of the Methodology section. Research datasets produced by the team are published in data-focused journals—such as *Scientific Data*, *Data in Brief*, *Data*, and *Data Intelligence*—or archived in open repositories like Zenodo and OSF to ensure broad accessibility and improve data reusability within the scientific community.

In coaching activities, ISR further encourages participants to embrace open science by actively using secondary datasets from data journals, public data-sharing platforms, and reputable repositories such as Mendeley Data, Harvard Dataverse, OSF, Zenodo, and the Science Data Bank. To cultivate transparency, the research team also uploads complete manuscripts to preprint servers and disseminates them within the SM3D community, accompanied by detailed information on methodology, scripts, and data.

Third, enhancing data-handling efficiency is expected to generate a form of "time surplus" relative to pre-tool workflows. This surplus can then be reinvested in deeper problem formulation and more rigorous interpretation of results. To support this objective, BMF analytics was designed to enhance researchers' creativity and productivity. Unlike earlier Bayesian analysis tools—such as WinBUGS, OpenBUGS, or JAGS—bayesvl allows researchers to construct models using relationship trees rather than using mathematical equations directly. For researchers, discovering novel and high-value knowledge requires substantial time devoted to thinking and imagination. Relationship trees increase productivity, flexibility, and intuitiveness in model construction, benefiting both novices and experienced experts (La et al., 2022).

Additionally, BMF analytics—particularly in its most recent extensions incorporating Granular Interaction Thinking Theory (GITT)—is designed to enable users to employ geometric concepts (e.g., points, lines, networks, circles) to conceptualize information interactions across hierarchical levels through mathematical thinking. Indeed, theoretical physicists such as Albert Einstein and Isaac Newton viewed geometry as a fundamental lens for understanding reality (Atiyah, 2005; Brooks, 2025). The application of geometry in BMF analytics' conceptual reasoning is illustrated, for example, in recent studies on Bitcoin value formation through information interactions across social and technical entropy systems (see Figures 1 and 2).

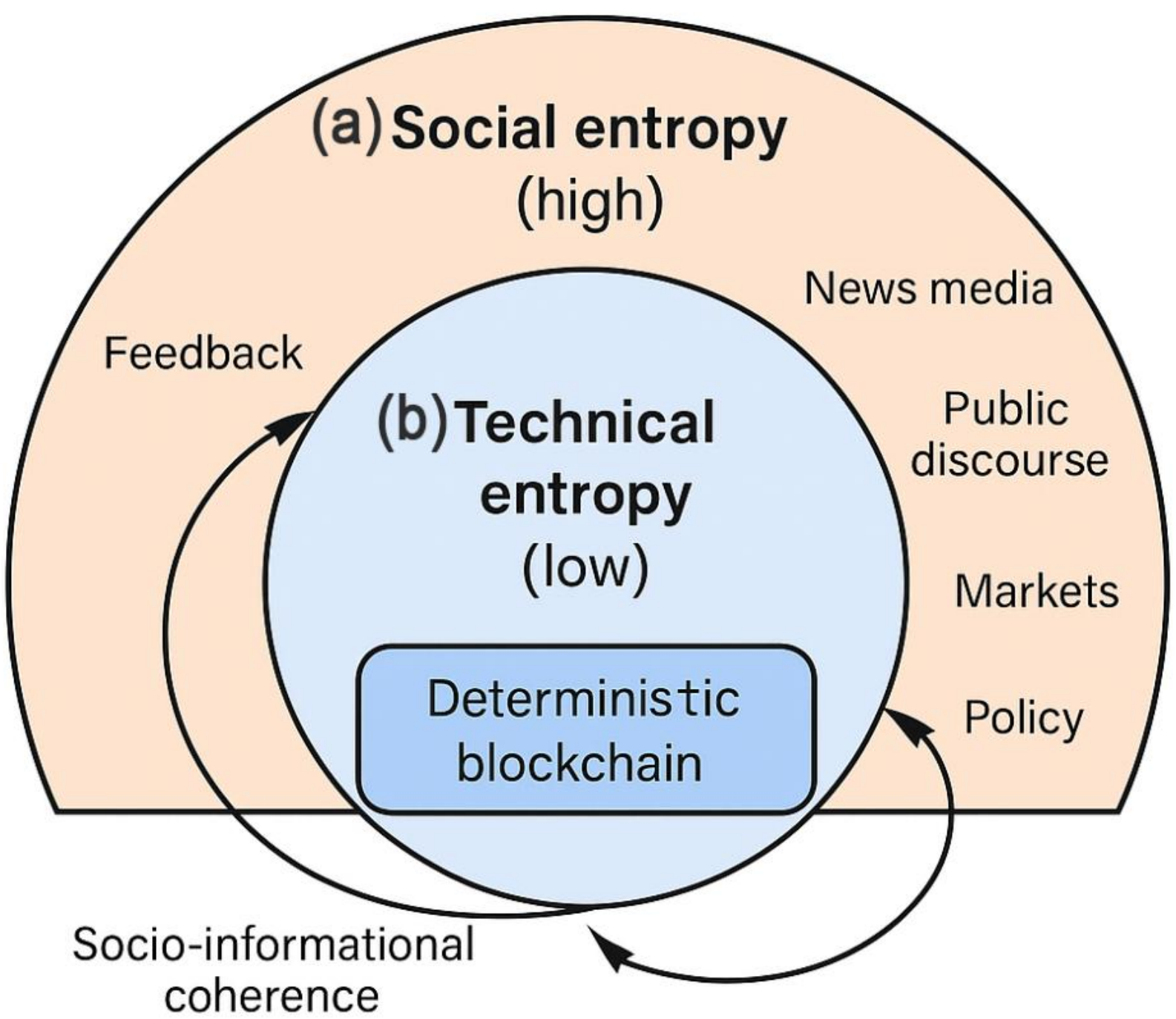

**Figure 1.** Bitcoin's entropy regulates the socio-technological ecosystem. Adapted from (Vuong, La, & Nguyen, 2025a)

Fourth, the team does not consider editorial or peer-review standards as immutable "golden rules," but instead regards rigorous investment in logical consistency, theoretical grounding, and the coherence of relational arguments as core values.

Fifth, the research team is not hesitant to continue innovations—whether in software, methodology, or theory—when such innovations are deemed necessary. These two principles have continually oriented the team toward improving the quality and robustness of the method. Since the inception of the bayesvl package, BMF analytics has undergone five major transformative innovations:

(i) the creation of the bayesvl software;
(ii) its integration with the Mindsponge mechanism;
(iii) the development of BMF analytics scientific reporting standards;
(iv) the upgrade of bayesvl to version 1.0.0; and

(v) the integration of GITT and mathematical logic into theoretical reasoning.

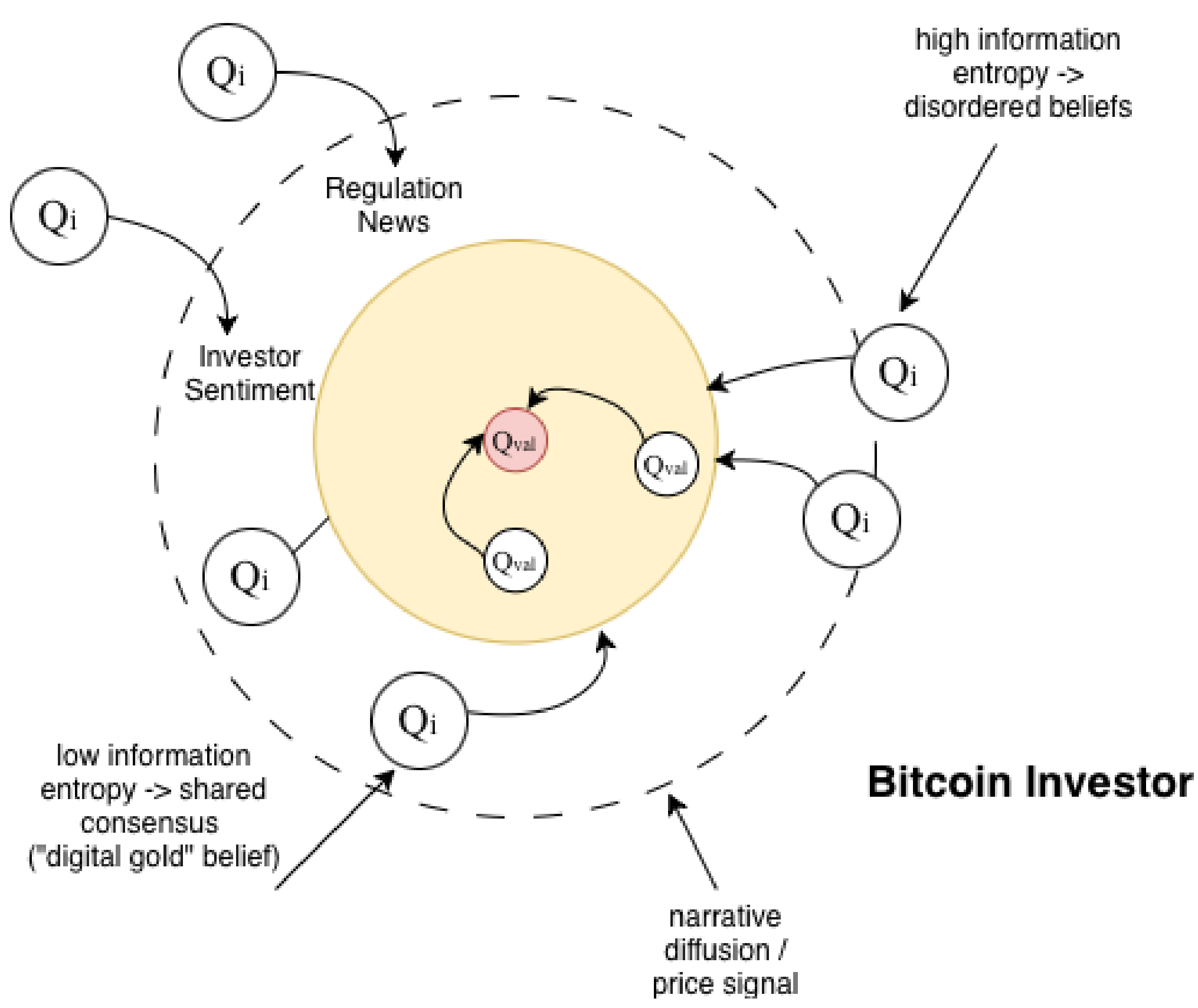

**Figure 2.** Information-value interaction diagram. Retrieved from (Vuong et al., 2025a)

Based on these principles, the knowledge required for methodological development is partitioned, connected, and organized according to priority and relevance to the method's objectives, philosophies, and principles. Knowledge that is less directly relevant or of lower immediate importance is retained for reference (assigned a low-probability status), while knowledge that is entirely irrelevant or excessively costly is filtered out.

### 3.3. Validation, Refinement, and the Discipline of Methodological Innovation

#### *3.3.1. bayesvl R package*

Initially, the idea of a new method emerges within the minds of researchers through interactions between their existing knowledge and observations from the surrounding environment—including working experiences, peer interactions, and engagement with scholarly literature. These interactions stimulate the search for solutions that can address concrete survival and development demands. Once a methodological idea takes shape, it must be tested, refined, and further developed through iterative interactions with diverse domains of knowledge.

Submission to CRAN also entails external expert evaluation and validation. After more than 18 months of development—comprising over 3,000 lines of code—bayesvl entered its final refinement phase between May 6 and May 17, 2019. On May 18, 2019, bayesvl (v0.8.1) was officially submitted to the Comprehensive R Archive Network (CRAN). CRAN serves as the central software repository for the R programming environment, hosting current and previous versions of R, documentation, and contributed packages. Each submission is rigorously reviewed by the CRAN team, primarily composed of experienced software engineers. Following five rounds of revisions, reviewer feedback, and resubmissions, bayesvl (v0.8.5) was officially released on CRAN on May 24, 2019.

Beyond technical validation, the method continued to be refined through its application in empirical research. As BMF analytics was employed across studies, its strengths and limitations were progressively revealed through interactions with experts holding different experiences, knowledge bases, and worldviews. These interactions proved invaluable for further strengthening the method and addressing its shortcomings.

This phase constituted a prolonged and demanding process—not only in implementation but also in persuading the academic community, particularly journal editors and peer reviewers. The challenges were substantial and persistent. A primary obstacle was skepticism from scholars unfamiliar with Bayesian statistics, who frequently raised questions such as: Why not use traditional methods? What advantages does this approach offer? Why not rely on established MCMC software such as rstanarm instead of a new tool like bayesvl? How should results be interpreted or compared with frequentist outcomes such as *p*-values?

These questions were not only difficult to answer but also required patience, precision, rigor, and readiness for further scrutiny. The same sets of questions often re-emerged across journals, editorial boards, and disciplinary contexts, sometimes requiring the entire explanatory process to restart from the beginning. In many cases, manuscripts were rejected simply because editors or reviewers were unwilling to engage with unfamiliar methods. For this reason, the validation, refinement, and development of methodological innovations must be understood as a highly disciplined and iterative process.

Importantly, this process was not solely a learning journey for ISR researchers; it also actively involved editors and reviewers. For example, during the peer-review process of one study (Vuong, Ho, et al., 2020), a reviewer conducted an exceptionally thorough evaluation across two rounds. While endorsing the logical foundations of the approach, the reviewer identified several areas requiring improvement, including the absence of specific functions and algorithms for comparing model selection performance in bayesvl v0.8.5. To meet these requirements, additional computations had to be performed using other software and integrated with bayesvl outputs. After these technical demands were satisfied, the reviewer requested further calculations under specific edge cases to assess reliability beyond the scenarios reported in the manuscript. Notably, the reviewer independently ran analyses using bayesvl and cross-validated the results using a different, more familiar software package. The reviewer then shared selected bayesvl outputs with the authors and communicated the validation results directly to the editorial board on their behalf. Such engagement not only tested the functional robustness of bayesvl but also significantly enhanced its credibility.

Subsequently, the functions required for computation, comparison, and validation during the review process of Vuong, Ho, et al. (2020) were incorporated into bayesvl v0.8.5. Along with other accumulated experiences, these additions became foundational components of later upgrades to versions v0.9.0 and v1.0.0. As technical standards were progressively met and the method was validated across multiple disciplines and knowledge domains, BMF analytics gradually gained reliability and broader acceptance within the scientific community.

#### *3.3.2. BMF analytics*

The development of BMF analytics also emerged from a process in which the method continuously interacted with different streams of knowledge. Through this iterative engagement, the research team gradually recognized an unexpected compatibility between the Mindsponge mechanism and Bayesian inference. In this context, the Mindsponge mechanism was identified as possessing three properties that are particularly useful for analyzing human psychology and behavior:

(i) The Mindsponge process is simultaneously shaped by information from the external environment and by internal evaluative mechanisms—namely, cost–benefit assessments embedded within individuals' belief systems and mindsets.
(ii) Causality within the Mindsponge framework is inherently nonlinear due to the multi-layered and multidimensional nature of information processing.
(iii) Mindsponge operates as a dynamically updating process, in which beliefs and attitudes are constantly revised through inductive learning from new information.

Although these three characteristics provide a compelling and coherent framework for explaining psychological and behavioral phenomena, relatively few statistical methods are capable of fully operationalizing the Mindsponge approach. Bayesian statistics, however, emerged as a particularly well-suited analytical tool for realizing this potential in practice.

First, Bayesian statistics—supported by MCMC algorithms—offer a high degree of flexibility in model construction. This flexibility enables the implementation of a wide range of models, including hierarchical models, nonlinear models, and combinations of both, thereby allowing the first two characteristics of the Mindsponge mechanism to be effectively incorporated into empirical analysis.

Second, Bayesian statistics treat all quantities as probability distributions, allowing researchers to focus on specific parameters of interest without the need to exhaustively control for all others. Given the dynamic, complex, and continuously evolving nature of Mindsponge processes, the number of parameters requiring control in traditional statistical frameworks can become prohibitively large, increasing research costs and reducing feasibility. Moreover, as the number of control variables grows, so does the risk of questionable research practices—such as stargazing, *p*-hacking, or HARKing—where researchers exploit sample variability to produce seemingly favorable results.

Third, the belief-updating mechanism at the core of Bayesian inference provides a foundational alignment with the Mindsponge philosophy. Humans exist in an information-saturated and constantly changing world, and their psychological states and behaviors evolve in response to shifting contexts and personal needs. Consequently, many scientific findings may become outdated over time and fail to replicate. By enabling continuous updating as new evidence becomes available, Bayesian statistics offer an ideal methodological foundation for studying cognition and behavior through the Mindsponge lens.

Beyond these three core reasons, additional factors further reinforce the compatibility between Bayesian statistics and the Mindsponge mechanism—which later evolved into Mindsponge Theory and Granular Interaction Thinking Theory (GITT) (Vuong, 2023; Vuong & Nguyen, 2024b). Although additional points of compatibility have been identified over the years of applying BMF analytics, the three points outlined above constitute the foundational rationale for the initial integration of the Mindsponge mechanism and Bayesian inference.

Once BMF analytics had taken shape, its functionality continued to be validated through systematic application to open datasets and publicly available databases. Through this process, the method's potential to extract meaningful insights from datasets that were often perceived as having exhausted their analytical value has been recognized. At present, this capacity of BMF analytics has been demonstrated not only through studies conducted by the ISR research team but also through research outputs produced by members of the SM3D Community. These independent applications provide further evidence that BMF analytics is capable of uncovering latent informational value in existing data, thereby reinforcing its usefulness, adaptability, and relevance across diverse research contexts.

## GITT-VT analytical paradigm

In addition, the peer-review process across multiple disciplinary domains created diverse validation conditions for assessing the reasoning and modeling capabilities of BMF analytics. Exposure to heterogeneous epistemic standards, methodological expectations, and domain-specific questions subjected the framework to sustained analytical validation pressure. This process, in turn, created the conditions for the extension and further development of BMF analytics into the GITT–VT analytical paradigm, which is now used to examine the inception, refinement, and developmental trajectory of BMF analytics itself. The transition to the GITT–VT paradigm marks a substantial advance in strengthening the rigor and robustness of methodological reasoning and discussion, particularly through the systematic integration of mathematical logic, the worldview of quantum physics, the entropy, and geometric conceptualization.

In fact, integrating mathematical reasoning and physics knowledge into social sciences and humanities research had long been an aspiration of the ISR research team (Van Huu & Vuong, 2007; Van Huu, Vuong, & Ngoc, 2005; Vuong, 2001). However, this ambition was constrained by the absence of sufficiently powerful theoretical frameworks and adequately flexible statistical tools. This limitation remained unresolved until an unexpected theoretical advance occurred: the emergence of GITT. This advance arose while the team was seeking physics-based insights into the problem of shifting socio-economic and cultural value systems from eco-deficit to eco-surplus, a transformation aimed at addressing foundational shortcomings of mainstream economic thinking in solving the climate change and environmental degradation problems (Vuong & Nguyen, 2024a).

This advance was catalyzed by two serendipitous moments.

The first occurred when the researchers encountered *On the Origin of Time: Stephen Hawking's Final Theory* by Hertog (2023). In this book, Hertog (2023) argues that physical laws are not fixed but instead evolve with the universe over time—a mechanism that exhibits similarities to the core logic of Mindsponge Theory. Motivated by this insight, the research team began to explore quantum mechanics more deeply, benefitting from the works of Rovelli (2016, 2018).

Through this engagement, the researchers recognized profound parallels between Mindsponge Theory and the three defining characteristics of quantum mechanics: granularity, relationality, and indeterminacy. Specifically, within the Mindsponge Theory's view, the world is conceived as being constituted by informational "particles," interactions, and information-processing processes. Individuals, organizations, societies, and even living organisms can be understood as information-processing systems that seek to maximize perceived benefits and minimize perceived costs—including energetic costs—in pursuit of a fundamental objective: prolonging the system's existence. As external conditions change, these systems adjust their internal structures and processes in order to adapt and evolve. These characteristics are similar to granularity and relationality features in quantum mechanics.

Within the quantum mechanics worldview, the relationships among the fundamental constituents of space are commonly represented as networks of nodes and edges. This representational structure closely parallels the Bayesian networks employed in social analysis grounded in Mindsponge Theory. For instance, Figure 3-A depicts a graph consisting of nodes that represent elementary spatial grains and links that denote adjacent particles separated by surfaces, whereas Figure 3-B shows a Bayesian network derived from one of the BMF analytics-aided empirical studies. Furthermore, the BMF analytics—designed as a measurement apparatus for psychological and social information—conceptualizes all relationships as probabilistic rather than deterministic, thereby echoing the principle of indeterminacy that lies at the core of quantum physics.

Nevertheless, conceptual similarity alone was insufficient to bridge the worldview of quantum mechanics with the social sciences and humanities, or to resolve the central question raised in the book *Better Economics for the Earth* (2024a), concerning the transformation of value systems. Addressing this challenge required answering a more fundamental question: What is value, and does it have any connection to physical principles?

At this point, a second serendipitous moment occurred. The researchers unexpectedly encountered an answer in Carlo Rovelli's discussion of the relationship between heat, time, and information (Rovelli, 2018). This insight naturally led to Boltzmann's entropy and the associated concepts of information loss and energy interaction. Through this intellectual pathway, a critical connection emerged: an informational-entropy-based notion of value, which provides a conceptual bridge between physical processes and socio-economic value formation (Vuong et al., 2025b).

This realization laid the theoretical foundation for extending BMF analytics into the GITT–VT analytical paradigm, enabling a more rigorous, physically grounded, and logically coherent approach to studying value formation, transformation, and sustainability across social, economic, and ecological systems.

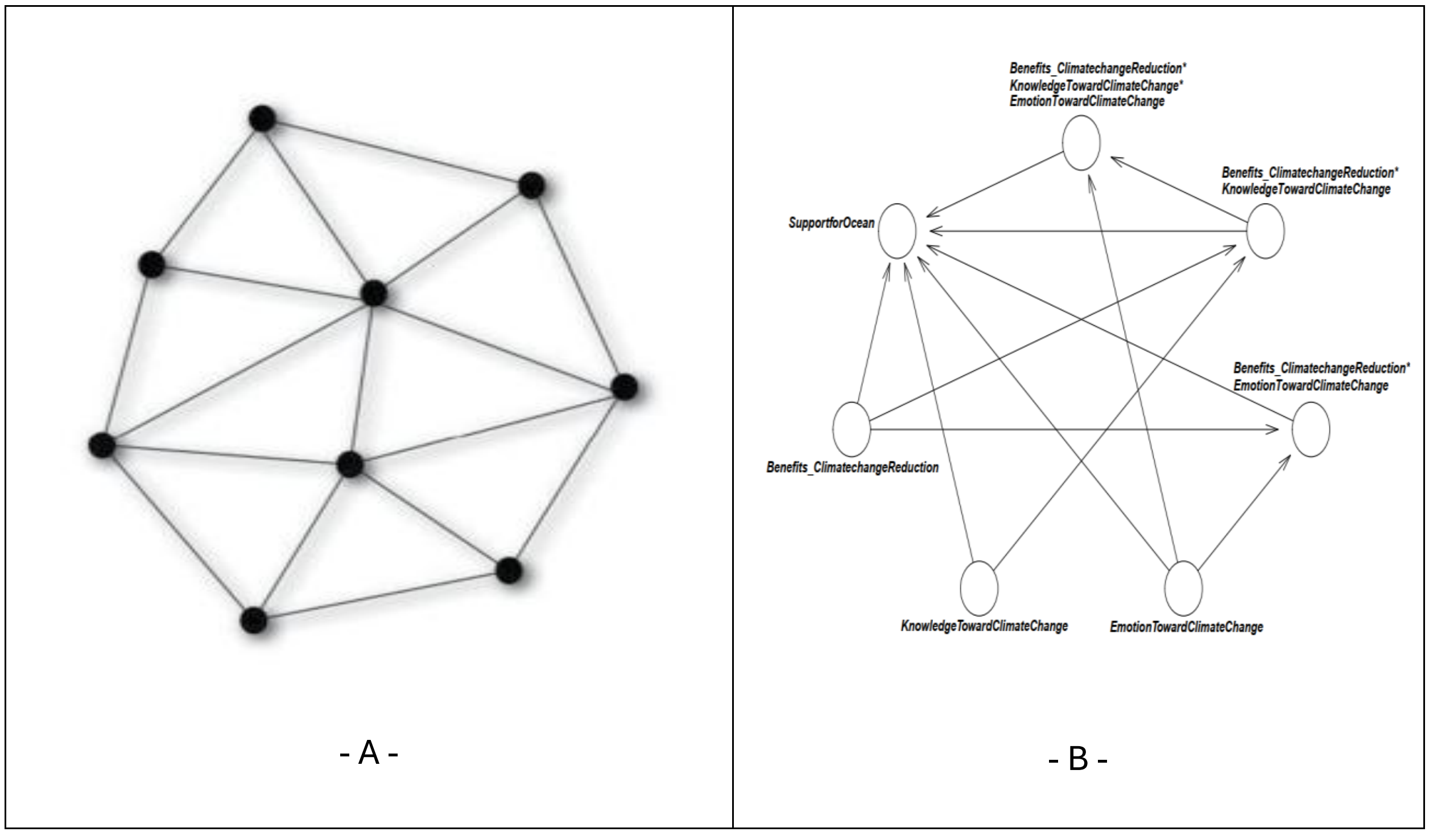

**Figure 3.** Left panel (A) depicts a graph representing the connections among elementary grains of space, while the right panel (B) presents a Bayesian network used in BMF analytics. (Adapted from Rovelli (2018) and Nguyen, Duong, Nguyen, La, and Vuong (2024))

### 3.4. The intergenerational knowledge transmission

Another crucial dimension in validating the usefulness of a method lies in its ability to be effectively operated by researchers other than its developers. To support ECRs in becoming familiar with Bayesian statistical tools and applying BMF analytics to their own research, the ISR research team initiated the establishment of the SM3D Community. Through this platform, researchers are able to participate in collaborative research projects alongside the developers of BMF analytics while receiving research training entirely free of charge. If a method is genuinely useful, participating members should gradually become capable of conducting independent research on their own and become new mentors.

This initiative emerged during the particularly challenging period of the COVID-19 pandemic, when the research team realized that even individuals with clear strategic direction, substantial experience, and high stress tolerance could still feel disoriented and trapped. This raised a critical question: How could other researchers—especially young scholars, those with limited resources, or those lacking access to formal methodological training—persist in an academic career under such conditions? This concern was compounded by intense competition within academia and mounting pressure from publication and citation metrics. In reality, many researchers withdrew from the academic path altogether. Even within ISR, several members left during the pandemic. These experiences became the primary motivation for founding the SM3D Community.

On June 22, 2022, after nearly three months of development beginning on April 7, 2022, the SM3D Community (SM3D Portal) officially commenced operation. The platform was designed to support researchers through on-the-job coaching in conducting research using BMF analytics, while simultaneously providing a space for knowledge exchange and community building. SM3D stands for Serendipity–Mindsponge–3D knowledge management framework (Jin et al., 2023).

The SM3D platform aims to train and enhance research, writing, and publication capacities for ECRs and researchers working under resource constraints, guided by three primary objectives:

(i) To promote global knowledge exchange and collaboration grounded in openness, transparency, equity, and innovation;
(ii) To create opportunities for researchers, particularly ECRs and those from low-resource regions, to participate in collaborative research projects applying BMF analytics;
(iii) To provide an enabling environment for participants to discuss ideas, produce scholarly outputs, and improve research skills, with theoretical, technical, and editorial support from the ISR research team.

To date, the SM3D Community has implemented over 100 collaborative research projects, training nearly 110 researchers from 67 institutions across 22 countries, with 85.2% of participants originating from developing countries. Partner institutions associated with SM3D participants include, *inter alia*, the University of Calcutta (India), China University of Political Science and Law (China), Monash University (Australia), the University of Pretoria and the University of the Western Cape (South Africa), Pepperdine University (USA), Sciences Po (France), Western University (Canada), Saint Louis College (Thailand), Widya Mandala Catholic University Surabaya (Indonesia), Ton Duc Thang University and Hanoi

University of Science and Technology (Vietnam), and Chinhoyi University of Technology (Zimbabwe).

To date, BMF analytics has demonstrated both effectiveness and operational viability through concrete scholarly outputs and the observable professional growth of participating members. Through the SM3D Community and its structured BMF analytics coaching program, a positive academic environment has been cultivated—one in which ECRs and researchers in resource-limited settings can identify and leverage development opportunities, thereby sustaining motivation and momentum. After three years of SM3D's operation, several members have successfully applied BMF analytics to their own survival and development needs, achieving notable academic milestones. Two illustrative cases are Dr. Minh-Phuong Thi Duong (Ton Duc Thang University, Vietnam) and Dr. Ni Putu Wulan Purnama Sari (Widya Mandala Catholic University Surabaya, Indonesia).

Dr. Duong joined the SM3D Community relatively early, around March 2023, while seeking effective methodological approaches to address research challenges and strengthen her academic capacity. Her first project, "BMF CP36: Predictors of people's support for a policy focus on marine and coastal preservation," marked her initial engagement. Despite early difficulties in absorbing unfamiliar concepts, strong community support and structured BMF analytics resources significantly helped ease her learning experiences. Her first scholarly output—examining stakeholder support for marine protection policies—was published in *Sustainability* on August 10, 2023 (Nguyen, Duong, et al., 2023).

Dr. Duong soon transitioned from participant to mentor, actively coordinating multiple projects within the community. Notably, she served as corresponding author for "Beyond trust: How usefulness and immersiveness drive space tourism intentions in high-risk contexts," published in *Current Issues in Tourism*, a leading journal in tourism and hospitality management (Nguyen et al., 2025). She has also coordinated other projects and mentored new members, with additional publications in peer-reviewed journals such as *Animal Production Science* (Australian Academy of Science) and *Sustainable Water Resources Management* (Springer) (Duong, Li, Nguyen, & Vuong, 2025; Duong, Sari, Mazenda, Nguyen, & Vuong, 2024).

Dr. Sari joined SM3D around the same time as Dr. Duong and quickly demonstrated remarkable initiative and intellectual openness by applying BMF analytics to ongoing collaborative research. These efforts soon yielded significant outcomes [88,90,100,107].

Beyond learning, she actively proposed and led collaborative projects, notably a research series on school nutrition. The first two studies in this series were published in the *Journal of Hunger & Environmental Nutrition* (Taylor & Francis) (Agung et al., 2025; Sari, Mazenda, Katiyatiya, Nguyen, & Vuong, 2025), and soon followed by the study in the *International Journal of Sociology and Social Policy* (Huni et al., 2025)

Through SM3D collaboration, Dr. Sari rapidly built an international research network spanning Vietnam, China, Australia, and several African countries. She was subsequently invited to mentor doctoral researchers at the University of Pretoria (South Africa), the University of Nairobi (Kenya), and Chinhoyi University of Technology (Zimbabwe) through the school nutrition research series. Less than two years earlier, she had been an ECR seeking methodological direction; she has since become a mentor herself within SM3D. In recognition of her academic contributions, she was nominated as an Outstanding Staff Member of Widya Mandala Catholic University Surabaya in 2024.

Cases such as Dr. Duong and Dr. Sari illustrate how, once methodological effectiveness is achieved, participants can move toward proactive program development, sometimes at scales exceeding initial expectations. A notable example is Dr. Sari's mentorship within the PhD training program at the University of Pretoria, which has contributed to the emergence of a third-generation (F3) cohort of BMF analytics users.

From the initial developers (F0) to core method builders (F1), and now to expanding third- and fourth-generation users (F3 and beyond), a relay-like process of knowledge transmission has taken shape. Although simplified in schematic representations, the actual scale of this diffusion is larger.

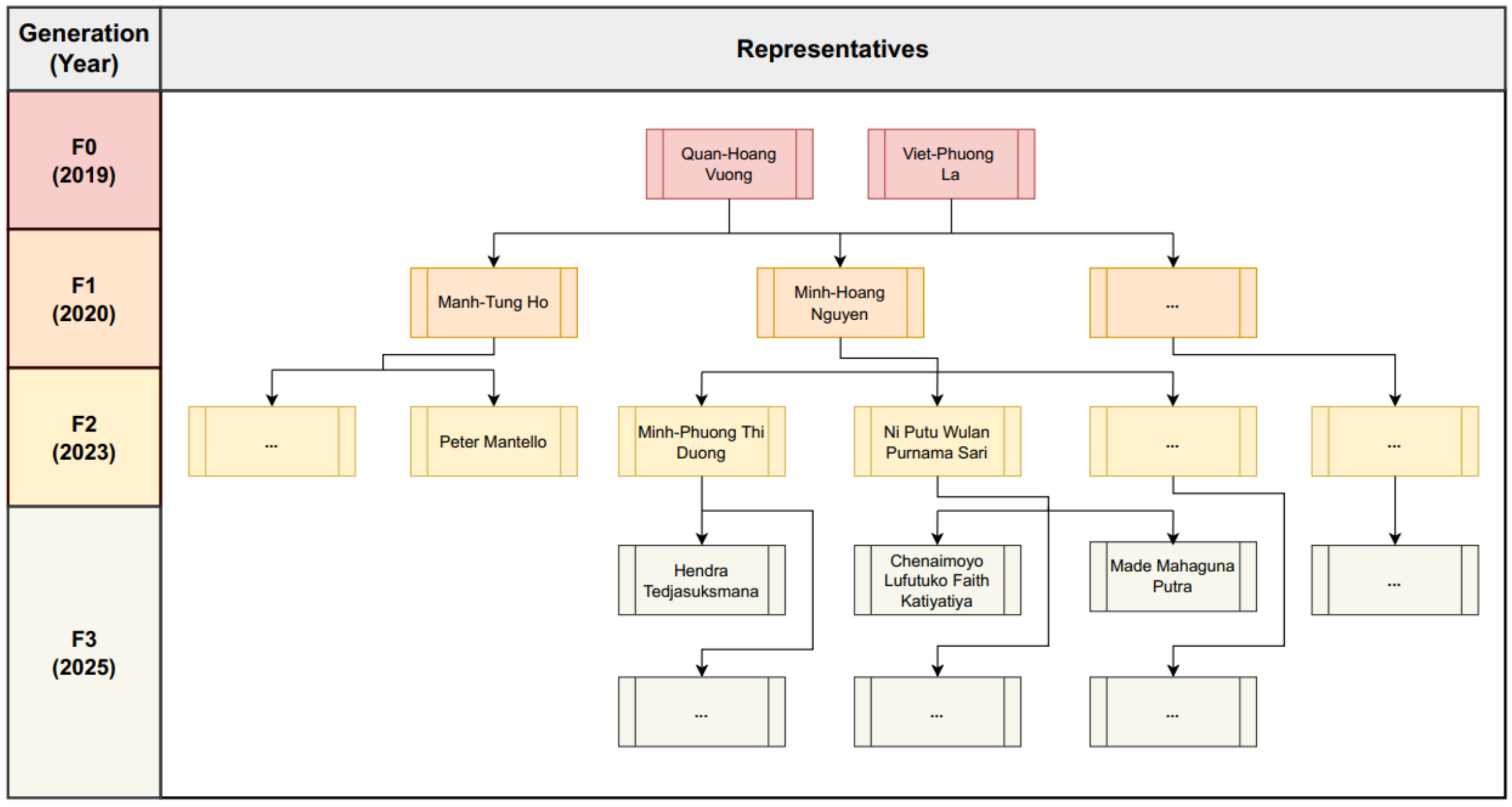

**Figure 4.** The generational tree of BMF analytics users

This intergenerational relay marks a transition from merely meeting minimum requirements or satisfying curiosity toward researchers independently designing long-term academic programs. As a result, a growing number of participants are now prepared to assume mentoring roles, gradually alleviating the initially severe shortage of mentors and reinforcing the sustainability of the BMF analytics ecosystem.

## 4. Final remarks

Through the case study of BMF analytics, it becomes evident that methodological innovations often emerge from survival-driven needs and continue to evolve in order to sustain that very survival over time. In its early stages, BMF analytics was developed as a pragmatic response to concrete constraints—limited resources, high uncertainty, and the need to maintain scientific productivity under adverse conditions. As these constraints changed, the method itself had to be continuously refined to remain viable, reliable, and useful. In this sense, methodological development is not a one-off technical achievement but an adaptive and cumulative process. The long-term sustainability of a methodology does not rest solely on its original design or its initial developers, but also depends on whether subsequent generations of researchers perceive the method as genuinely useful for addressing their own research problems and survival pressures within evolving academic environments. When users experience tangible benefits—such as reduced costs, enhanced analytical clarity, or increased research autonomy—they will become motivated not only to adopt the method but also to improve, extend, and reinterpret it.

When development of the method first began, the bayesvl program was conceived with modest and deliberately constrained ambitions. Its overarching objective was to achieve flexibility in handling data types that are predominantly discrete variables. This design choice reflected practical considerations: the research team operated under limited resources, lacking both the personnel and financial capacity to outsource development. As a result, careful attention had to be paid to balancing functionality against available resources. Experience suggested that increasing functional capacity by as little as 20% could potentially require two to three times more development and testing time—or even longer—while simultaneously reducing the team's ability to control and debug errors. Given these trade-offs, ISR intentionally chose to maintain a restrained level of functionality that would nevertheless be sufficient for the majority of discrete-data applications commonly encountered in the social sciences and humanities.

This strategic restraint also left substantial room for future methodological enhancement. Subsequent generations of researchers can continue to build upon the existing framework, extending the program to accommodate more complex data structures—such as time-series data—and to integrate analytical functions beyond multilevel modeling. In this way, bayesvl in particular, and BMF analytics in general, was designed not as a closed or exhaustive solution, but as a stable and extensible foundation for ongoing methodological innovation.

Accordingly, the continued development and maturation of BMF analytics will be shaped by future researchers who internalize its principles, apply it across new domains, and actively contribute to its refinement. In this way, the survival logic that originally gave rise to the method is transformed into a collective, intergenerational process, whereby methodological innovation persists through shared usefulness, adaptive learning, and ongoing co-development rather than through static preservation or top-down enforcement.

Final words are reserved for the spirit of *Libre Examen*. We trust that good analytical apparatuses need to aid investigators, including ECRs, effectively to the extent that they will have more time for thinking. To this end, this evaluation appears to have justified the functioning of the BMF analytics. And for us, who have been directly engaged in mentoring, working on manuscripts, and finishing the publishing obligations, we even had the privilege of figuring out the deep meaning of the Six Persimmons painting by Master Muqi in the 13th century (see Figure 5), using the very analytical approach discussed in this paper (Nguyen, 2025).

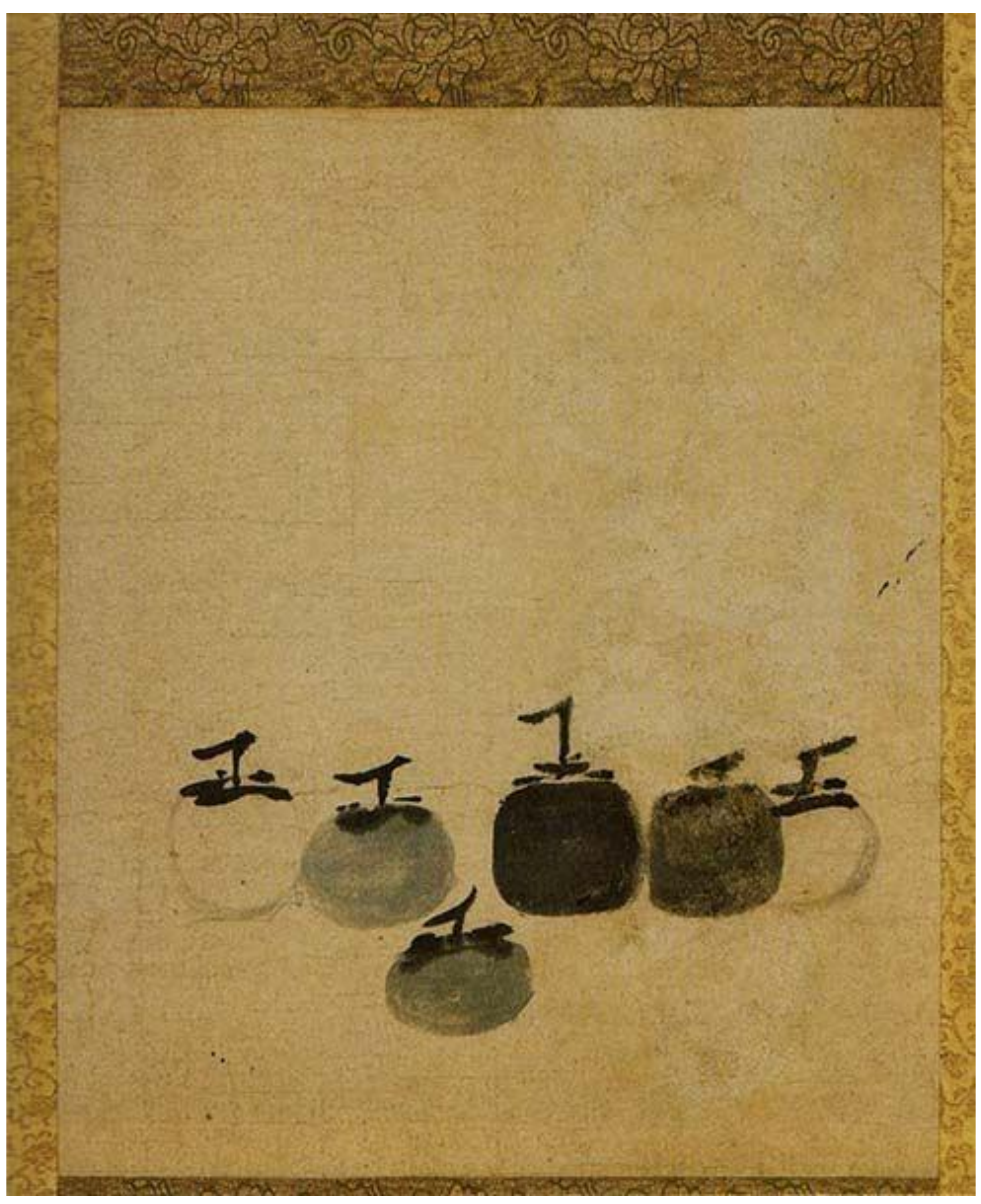

**Figure 5.** Six Persimmons (六柿圖), Muqi Fachang (1210–1269)